\documentstyle[12pt]{article}

\date{}
\begin{document}

\title{Generalization of Integrality Condition of
Prequantization to Phase Space with Boundaries\thanks{%
this work is supported by NSF of China, Pan Den Plan of China and LWTZ -1298
of Chinese Academy of Sciences}}
\author{{\ Ming-Xue Shao\thanks{{\protect\small E-mail: shaomx@itp.ac.cn}},
Zhong-Yuan Zhu\thanks{{\protect\small E-mail: zzy@itp.ac.cn}}} \\
{\small CCAST(World Laboratory), P.O.Box 8730, Beijing, 100080, P.R.China}\\
{\small Institute of Theoretical Physics, Academia Sinica, P.O.Box 2735,
Beijing 100080, P.R.China.}}
\maketitle

\begin{abstract}

The Weil's integrality condition of prequantization line bundle is generalized 
to phase space with boundaries. The proofs of both necessity and  sufficiency 
are
 given. It is pointed out via the method of topological current that
Weil's integrality condition  is closely connected with the summation of index 
of 
isolated singular points of sections
of prequantization line bundle.  \\

\end{abstract}

\vskip 0.6in

Symplectic geometrical description of classical mechanics and its geometric
quantization are essentially globalization of, respectively, Hamiltonian
mechanics and canonical quantization\cite{Wood Sniatycki}. Geometric 
quantization has 
been
considered as a so far most mathematically thorough approach to
quantization. 
The first step in geometric quantization is prequantization
which needs a Hermitian line bundle to realize the representation of 
the Poission brackets of classical observables. Furthermore in
order to satisfy the Dirac condition\cite{dirac20} in prequantization, the pull 
back 
of curvature form $\Omega $ of the bundle should be the same as the symplectic 
form 
$\hbar ^{-1}\omega $ of the
symplectic manifold $M.$ The question is whether or not any $\omega $ can be 
used
to construct the bundle .  The answer is as follows. Such a bundle and
connection exist if and only if $\omega $ satisfies Weil$^{\prime }$s
integrality condition\cite{weil}\cite{kostant23}:~~`The integral of $\omega $
over any closed oriented 2-surface in $M$ is an integral multiple of $2\pi
\hbar .$' This condition is closely related to the quantization rule in the
old quantum theory\cite{Messiah}\cite{simms23}. In this paper we use the 
technique of
decomposition of connection \cite{duan1}\cite{smxmaster} and
topological current\cite{duan2}\cite{meng} to prove its necessity. We also
generalize it to the case of phase spaces with boundary. Besides a proof of
sufficiency via geometric construction is  given when $M$ is simply
connected.

Let $B \mapsto M$ be a Hermitian line bundle with $M$   the phase space of 
classical 
system and $\Psi =\psi s$ be the section of  $B$  with $s$
the unit section. The Dirac condition in prequantization requires\cite{Wood 
Sniatycki}
\begin{equation}
\Omega =\frac 1\hbar {\omega }  ,\label{2217}
\end{equation}
in which $\Omega$ is the curvature form  of the bundle $B$ and $\omega$ the 
symplectic 
form of symplectic manifold $M.$  The connection  on $B$ is defined as
\begin{equation}
D s=-i\Theta s, 
\label{230}  
\end{equation}
where  $\Theta$ denotes the connection 1-form. 
The covariant derivative of $\Psi $ is 
\begin{equation}
D \psi s=(d\psi -i\Theta \psi )s  
\label{231}
\end{equation}
which can be rewritten as 
\begin{equation}
D\psi =d\psi -i\Theta \psi ,  \label{232}
\end{equation}
from which we can obtain 
\begin{equation}
\Theta =-i\frac 1\psi d\psi +i\frac 1\psi D\psi .  \label{233}
\end{equation}

Now we write $\psi $ by its real part $\psi _1$ and imaginary part $\psi _2$ 
\begin{equation}
\psi =||\psi ||(n^1+in^2),  \label{235}
\end{equation}
where 
\begin{equation}
||\psi ||=\sqrt{{\psi _1}^2+{\psi _2^2}} ; \label{236}
\end{equation}
\begin{equation}
n^1=\frac{\psi _1}{\sqrt{{\psi _1}^2+{\psi _2^2}}},~~~~~n^2=\frac{\psi _2}{%
\sqrt{{\psi _1}^2+{\psi _2^2}}}  . \label{237}
\end{equation}
Denote $\vec{\psi}=(\psi _1,\psi _2)$ which can be considered as a
2-component vector field and $\vec{n}=(n^1,n^2)$ a unit vector field

\begin{equation}
{n^1}{n^1}+{n^2}{n^2}=1.  \label{238}
\end{equation}
Putting (\ref{235}) into (\ref{232}), we obtain 
\begin{equation}
D\psi=d||\psi||(n^1+in^2)+||\psi||(Dn^1+iDn^2),  \label{2316c}
\end{equation}
in which 
\begin{equation}
Dn^1=dn^1+\Theta n^2;~~~~~~Dn^2=dn^2-\Theta n^1 .  \label{2316a}
\end{equation}
Recall the $so(2)$ covariant derivative 
\begin{equation}
Dn^a=dn^a+\omega^{ab}n^b  . \label{2316b}
\end{equation}
The comparison of (\ref{2316a}) and (\ref{2316b}) implies $%
\Theta=\omega^{12} $, for which the underlined reason is the homomorphism of
Lie algebra of $su(1)$ and $so(2)$.

Putting Eqs. (\ref{2316c}) and (\ref{235}) into (\ref{233}), we get 
\begin{equation}
\Theta =\epsilon _{ab}n^adn^b-\epsilon _{ab}n^aDn^b , \label{239}
\end{equation}
where we have used Eq. (\ref{238}). From Eq. (\ref{239}), the curvature of
the line bundle is 
\begin{equation}
\Omega =d\Theta =\epsilon _{ab}dn^a\wedge dn^b-d(\epsilon _{ab}n^aDn^b)
.
\label{2310}
\end{equation}
If there is no zero points of $\psi $, or no singular points of $n,$ the
integral of (\ref{2310}) over any closed 2-surface will vanish by Stocks
theorem.. However in general (\ref{2310}) can not be globally correct if $%
\psi $ as a global section has zero points which depend on topological
property of the bundle. That is to say, (\ref{2310}) has singular points. In
this case a convenient method is to use the so called topological current
technique\cite{duan2} \cite{duan1}\cite{meng}as follows. The first term of
rhs of (\ref{2310}) is 
\begin{equation}
\epsilon _{ab}dn^a\wedge dn^b=2\pi Tdx^1\wedge dx^2
\end{equation}
where $x^\mu ,\mu =1,2$ is two dimensional local coordinates of oriented
2-surface $\Sigma $ in $M$, and $T$ is defined by 
\begin{equation}
T=\frac 1{2\pi }\epsilon ^{\mu \nu }\epsilon _{ab}\partial _\mu n^a\partial
_\nu n^b.  \label{2311}
\end{equation}
Substituting (\ref{237}) into (\ref{2311}) and using 
\begin{equation}
\partial _\mu n^a=\frac{\delta ^{al}||\psi ||^2-\psi ^a\psi ^l}{||\psi ||^3}%
\partial _\mu \psi ^l  \label{2311a}
\end{equation}
and 
\begin{equation}
\frac \partial {\partial \psi ^l}(ln\frac 1{||\psi ||})=-\frac{\psi ^l}{%
||\psi ||^2}  \label{2311b}
\end{equation}
results in 
\begin{equation}
T=-\frac 1{2\pi }\epsilon ^{\mu \nu }\epsilon _{ab}\partial _\mu \psi
^l\partial _\nu \psi ^b\frac \partial {\partial \psi ^l}\frac \partial
{\partial \psi ^a}(ln\frac 1{||\psi ||})  . \label{2311c}
\end{equation}
Define Jacobian determinant as 
\begin{equation}
\epsilon ^{ab}J([\frac{\partial \psi }{\partial x}])=\epsilon ^{\mu \nu
}\partial _\mu \psi ^a\partial _\nu \psi ^b,  \label{2311d}
\end{equation}
by virtue of the 2-dimensional Laplacian relation\cite{laplacian} 
\begin{equation}
\frac \partial {\partial \psi ^l}\frac \partial {\partial \psi ^l}(ln\frac
1{||\psi ||})=-2\pi \delta (\vec{\psi}) , \label{2311e}
\end{equation}
in which $(\frac{\partial ^2}{\partial \psi ^l\partial \psi ^l})$ is
2-dimensional Laplacian operator in $\vec{\psi}$ space, the $\delta $%
-function-like density 
\begin{equation}
T=\delta (\vec{\psi})J([\frac{\partial \psi }{\partial x}])  \label{2311f}
\end{equation}
is obtained. Suppose that the function $\psi ^a(a=1,2)$ possess $n$ isolated
zeroes. Let the i-th zero be $\vec{x}=\vec{z_i}$, one has 
\begin{equation}
\psi ^a(\vec{z_i})=0,with~~i=1,2,...,n.  \label{2311g}
\end{equation}
Then we have 
\begin{equation}
\delta (\vec{\psi})=\sum\limits_{i=1}^n\frac{\beta _i}{|J([\frac{\partial
\psi }{\partial x}])|_{\vec{x}=\vec{z_i}}}  , \label{2311h}
\end{equation}
where $\beta _i$ is a positive integer called Hopf index\cite{hopf} of the
i-th singular point, which denote the times the function $\vec{\psi}$ covers
the corresponding region while the point $\vec{x}$ covers the neighborhood
of $\vec{x}=\vec{z_i}$ once. Substituting (\ref{2311h}) into (\ref{2311f}),
the charge density $T$ can be written in the form

\begin{equation}
T=\sum\limits_{i=1}^n\beta _i\eta _i\delta ^2(\vec{x}-\vec{z_i}) ,
\label{2312}
\end{equation}
where 
\begin{equation}
\eta _i=\frac{J([\frac{\partial \psi }{\partial x}])}{|J([\frac{\partial
\psi }{\partial x}])|}|_{\vec{x}=\vec{z}_i}=\mp 1  \label{2312a}
\end{equation}
are called the Brouwer degrees\cite{brouwer}, which reflects whether or not
the covering of $\psi $ has the same direction on bundle with that of $\vec{x%
}$ on 2-dimensional base manifold.

Now consider the integral of (\ref{2310}) over a closed 2-surface $\Sigma $.
. Using (\ref{2312}) and the fact $\Sigma $ is closed, we get 
\begin{equation}
\int_\Sigma \Omega =2\pi \sum\limits_{i=1}^n\beta _i\eta _i=2\pi g
 ,
\label{2313}
\end{equation}
where $g=\sum\limits_{i=1}^n\beta _i\eta _i$ is topological charge.
Considering (\ref{2217}), 
\begin{equation}
\int_\Sigma \omega =hg.  \label{2313a}
\end{equation}
is obtained. This demonstrates the integral of $\omega $ over any closed
2-surface of phase space is an integer multiplied by plank constant $h$.
Furthermore from this derivation we know this integer can be determined by
the summation of index of zero-points of $\psi $. Now we consider the case
that 2-surface $\Sigma $ with boundary $\partial \Sigma $ to be piece-wise
smooth. Making use of Stocks theorem, we have 
\begin{equation}
\int_\Sigma \Omega =2\pi \sum\limits_{i=1}^n\beta _i\eta _i+\int_{\partial
(\Sigma -\Sigma ^{\prime })}\epsilon _{ab}n^adn^b-\int_{\partial \Sigma
}\epsilon _{ab}n^aDn^b  , \label{2314}
\end{equation}
where $\Sigma ^{\prime }$ is another 2-surface with boundary $\partial
\Sigma ^{\prime }$ which is chosen to be smooth and $\rightarrow \partial
\Sigma $ in limit. Defining $\alpha $ to be the angle from $\vec{e_1}=(1,0)$
to $\vec{n}=(n^1,n^2)$, we get 
\begin{equation}
\epsilon _{ab}n^adn^b=d\alpha  . \label{2315}
\end{equation}
Obviously the change of $\alpha $ along $\partial \Sigma ^{\prime }$ is $%
2\pi l$ with $l$ an integer. So, considering this and (\ref{2217})(\ref{2314}%
), we obtain 
\begin{equation}
\int_{ \Sigma }\omega =2\pi \hbar (-l+\sum\limits_{i=1}^n\beta
_i\eta _i)+\hbar \int_{\partial \Sigma }\Theta ,  \label{2316}
\end{equation}
in which we have used the Stocks theorem and (\ref{2217})(\ref{2314}).
Notice the second term in (\ref{2316}) is invariant under the transformation
of connection. Condition (\ref{2316})is necessary. A slight difference is
that here we generalize to 2-surface with boundary in $M$. When $M$ has
boundary this generalization is needed because a closed path $\gamma $ in $M$
will deduce two types of 2-surfaces in $M$ : (1) with $\gamma $ as boundary
(2) with $\gamma $ and a closed path $c\subset \partial M$ as boundary. The
particular case is when $M$ is closed both the second term in (\ref{2316})
and $l$ in the first term vanish, we return 
to
\begin{equation}
\int_{ \Sigma }\omega =2\pi \hbar (\sum\limits_{i=1}^n\beta _i\eta
_i).  \label{2322}
\end{equation}

For a general case that the phase space $M$ has boundary $\partial M$,
assuming that $M$ is simply connected, (\ref{2316}) is also a sufficient
condition. We will use geometric construction to prove it.

Suppose that $\Omega =\omega /{\hbar }=d\Theta $ is a closed 2-form with $%
\omega $ satisfy (\ref{2316}). Choose a base point $m_0$ in $M$ and let $K$
denote the set of all triples $(m,z,\gamma )$ where $m\in M$, $z\in {\cal C}$%
, and $\gamma $ is a piecewise smooth path from $m_0$ to $m$. On $K$, define
an equivalence relations $\sim $ by

\begin{equation}
(m,z,\gamma )\sim (m^{\prime },z^{\prime },\gamma ^{\prime })  , \label{2323}
\end{equation}
whenever $m=m^{\prime }$ and 
\begin{equation}
z^{\prime }=\{ 
\begin{array}{l}
zexp(i\int_\Sigma \Omega ), \\ 
zexp(-i\int_{\Sigma ^{\prime }}\Omega +i\int_c\Theta ),
\end{array}
\label{2324}
\end{equation}
where $\Sigma $ is any surface with boundary made up of $\gamma (\gamma
^{\prime })^{-1}$ and $\Sigma ^{\prime }$ surface with boundary made up of $%
\gamma ^{\prime }(\gamma )^{-1}c$ with $c\subset \partial M$ a closed path.
Because (\ref{2316}) hold, it does not matter which surface is chosen.

We shall take as the total space of our bundle the manifold $B=K/\sim $ with
the obvious projection onto $M$: addition and scalar multiplication within
the fibers are defined by 
\begin{equation}
\lbrack (m,z,\gamma )+(m,z^{\prime },\gamma )]=[(m,z+z^{\prime },\gamma )]
 ,
\label{2325}
\end{equation}
\begin{equation}
w[(m,z,\gamma )]=[(m,wz,\gamma )];w\in C  \label{2326}
\end{equation}
(where square brackets denote equivalence classes). The local
trivializations are constructed as follows: Let $m_1\in m$ and $U_1$ be a
simply connected neighbourhood ( possibly containing points of boundary $%
\partial M$) of $m_1$ on which there is a real 1-form $\Theta _1$ such that $%
d\Theta _1=\Omega $. Let $m\in U_1$ and let $\xi $ be a smooth curve in $U$
from $m_1$ to $m$. For each $w\in C$, put 
\begin{equation}
\psi _1(m,w)=[(m,wexp(-i\int_\xi \Theta ),\gamma \xi ^{-1})]  , \label{2327}
\end{equation}
where $\gamma $ is some fixed curve joining $m_0$ to $m_1$. It follows from
Stokes theorem and (\ref{2323}) that $\psi _1(m,w)$ is independent of the
choice made for $\xi $. Therefore $U_1$ and $\psi _1$ from a local
trivialization.

If $(U_2, \psi_2)$ is another such local trivialization by replacing $m_1,
U_1,$ and $\Theta_1$ by $m_2, U_2,$ and $\Theta_2$, and if $U_1\cap U_2$ is
simply connected, then

\begin{equation}
\psi_2(m, w)=c_{12}(m)\psi_1(m, w); m\in U_1\cap U_2, w\in C  , \label{2328}
\end{equation}
where the transition function $c_{12}\in C_{C}^{\infty}(U_1\cap U_2)$
satisfies

\begin{equation}
c_{12}(m^{\prime })=c_{12}(m)exp(i\int_m^{m^{\prime }}(\Theta _1-\Theta
_2));m,m^{\prime }\in U_1\cap U_2  , \label{2329}
\end{equation}
in which the integral is taken along any path from $m$ to $m^{\prime }$. It
follows that 
\begin{equation}
\frac{dc_{12}}{c_{12}}=i(\Theta _1-\Theta _2)  , \label{2330}
\end{equation}
hence $\Theta $s are the connection 1-form to define covariant derivative $D$
on the bundle $B$ with curvature $\Omega $. Further, a compatible Hermitian
structure on $B$ is defined by 
\begin{equation}
([m,z,\gamma ],[m,z,\gamma ])=z{\overline{z}}.  \label{2331}
\end{equation}
Since a different base point will give an equivalent Hermitian line
bundle-with connection, we complete the proof of sufficient condition.

At last of this paper, we point out that the second term in (\ref{2316}) can be 
simplified
by choosing unit vector $\vec{n}$ to be tangent to the boundary. Then $%
n^b,k^b=-\epsilon _{ab}n^a$ has the same orientation with the base manifold,
then define 
\begin{equation}
-\epsilon _{ab}Dn^an^b=k_gds  \label{2332}
\end{equation}
with $k_g$ the geodesic curvature along the boundary and $s$ the parameter
of the boundary $\partial \Sigma $. Consider the boundary is piece-wise with 
$m$ angle-change points. Let the inner angle of i-th angle-change points be $%
\alpha _i$. From Eqs.(\ref{2316}) and (\ref{2332}), we get 
\begin{equation}
\int_{ \Sigma }\omega =2\pi \hbar \sum\limits_{i=1}^n\beta _i\eta
_i-\hbar \sum\limits_{i=1}^m(\pi -\alpha _i)-\hbar \int_{\partial \Sigma
}k_gds.  \label{2321}
\end{equation}

\end{document}